\documentclass[12pt]{iopart}

\usepackage{graphicx}
\begin{document}

\title[Phase Dependence of Phonon Tunneling in Bosonic S-I-S Junctions]{Phase Dependence of Phonon Tunneling\\
 in Bosonic Superfluid-Insulator-Superfluid Junctions}

\author{Ippei Danshita$^{1,2}$, Nobuhiko Yokoshi$^1$ and Susumu Kurihara$^1$}
\address{$^1$Department of Physics, Waseda University, Ohkubo, Shinjuku-ku, Tokyo, 169-8555, Japan}
\address{$^2$National Institute of Standards and Technology, Gaithersburg, MD 20899, USA}
\ead{danshita@nist.gov}

\begin{abstract}
We consider the tunneling of phonon excitations across a potential barrier spatially separating two condensates with different macroscopic phases. 
We analyze the relation between the phase difference $\varphi$ of the two condensates and the transmission coefficient {\it T} by solving the Bogoliubov equations. 
It is found that {\it T} strongly depends on $\varphi$, and that the perfect transmission of low-energy excitations disappears when the phase difference reaches the critical value which gives the maximum supercurrent of the condensate.
We also discuss the feasibility of observing the phase differences in experiments.
\end{abstract}

\maketitle

\section{Introduction}
Since the first observation of Bose-Einstein condensation (BEC) in trapped dilute atomic gases~\cite{rf:JILA, rf:MIT}, developments of experimental and theoretical studies in this field have advanced rapidly~\cite{rf:BEC, rf:leg}.
One of the greatest advantages of these systems is that experimentalists have useful tools to create various external potentials for atoms.
Magnetic traps, which are standard equipments to confine atomic gases, create harmonic potentials for atoms.
In early stages of the study, many characters of BECs in harmonic confinement were vigorously investigated.
The laser beams can also make periodic potentials for atoms, called optical lattices, by superposing the counter-propagating laser beams and preparing standing waves~\cite{rf:lattice}.
Moreover, the blue-detuned laser beams act as potential barrier for atoms, and one can create double-well potentials by focusing the laser beam into the center of a confining potential~\cite{rf:double}.
One can regard Bose-Einstein condensates in a double-well potential as a bosonic superfluid-insulator-superfluid (S-I-S) junction like a fermionic superconductor-insulator-superconductor junction.

The controllability of external potentials provides a suitable stage for study of quantum tunneling phenomena.
Actually, many theorists have investigated the Josephson-like effect, corresponding to a kind of quantum tunneling~\cite{rf:BEC, rf:leg, rf:smerzi}, since the experimental realization of BECs in a double-well trap~\cite{rf:double}.
Recently, some types of the Josephson effect, called Josephson plasma oscillation and self-trapping in Ref.~\cite{rf:smerzi}, were experimentally observed~\cite{rf:JPST}.

The study of elementary excitations is one of the main subjects to understand various properties of the Bose condensate, such as dynamics, thermodynamics and superfluidity.
Several interesting properties of the elementary excitations have been probed experimentally in gaseous BECs, including the observation of collective modes~\cite{rf:BEC,rf:dipo_jila,rf:dipo_mit} and the measurement of the Bogoliubov excitation spectrum with use of Bragg spectroscopy~\cite{rf:BEC,rf:bragg,rf:bogoex}.
As for the tunneling problem of the elementary excitations, it was predicted that low-energy excitations perfectly transmit through a potential barrier, and such a behavior was called the {\it anomalous tunneling}~\cite{rf:antun,rf:kov}.
A recent work shows that the anomalous tunneling property is crucial to the phononlike form of the excitation spectrum of BECs in a periodic potential, which is directly connected to the superfluidity~\cite{rf:kp}.

On the other hand, the macroscopic phase difference $\varphi$ plays a key role in Bose condensed systems. 
Experiments on the matter wave interference~\cite{rf:double} and the Josephson effect~\cite{rf:JPST} have clearly demonstrated the existence of the phase difference between two spatially-separated BECs. 
In addition, the problem of the tunneling of quasiparticles in the fermionic superconductor-insulator-superconductor junction has been studied in detail, and it is well known that the phase difference between two superconductors essentially affects the scattering processes at the potential barrier~\cite{rf:sis}.
It is also expected that the tunneling of the elementary excitations in bosonic S-I-S junctions depends on the phase difference.

In the present paper, we study the tunneling of the elementary excitations in a bosonic S-I-S junction and analyze the $\varphi$-dependence of the transmission coefficient $T$ by solving the Bogoliubov equations analytically.
It is found that {\it T} strongly depends on $\varphi$, with regard to the anomalous tunneling.
The peak width of the transmission coefficient decreases as $\varphi$ increases, and the peak vanishes when the phase difference reaches $\varphi_{\rm c}\simeq\frac{\pi}{2}$ which gives the maximum supercurrent of the condensate.
We will show that the anomalous tunneling and its remarkable $\varphi$-dependence originate from the existence of the components localized near the potential barrier.

The outline of the paper is as follows.
In Sec. II, we introduce a formulation of the problem using the mean field theory.
We calculate the condensate wave function and the relation between the supercurrent and the phase difference analytically.
In Sec. III, we analytically solve the Bogoliubov equations and obtain the transmission coefficient.
In Sec. IV, we discuss a mechanism of the anomalous tunneling.
We also discuss the feasibility of observing the anomalous tunneling in real systems.
We summarize our results in Sec. V.

\section{Mean field theory and model potential}
\begin{figure}[p]
\begin{center}
\vspace{60mm}
\includegraphics[width=4 in, height=2.5 in]{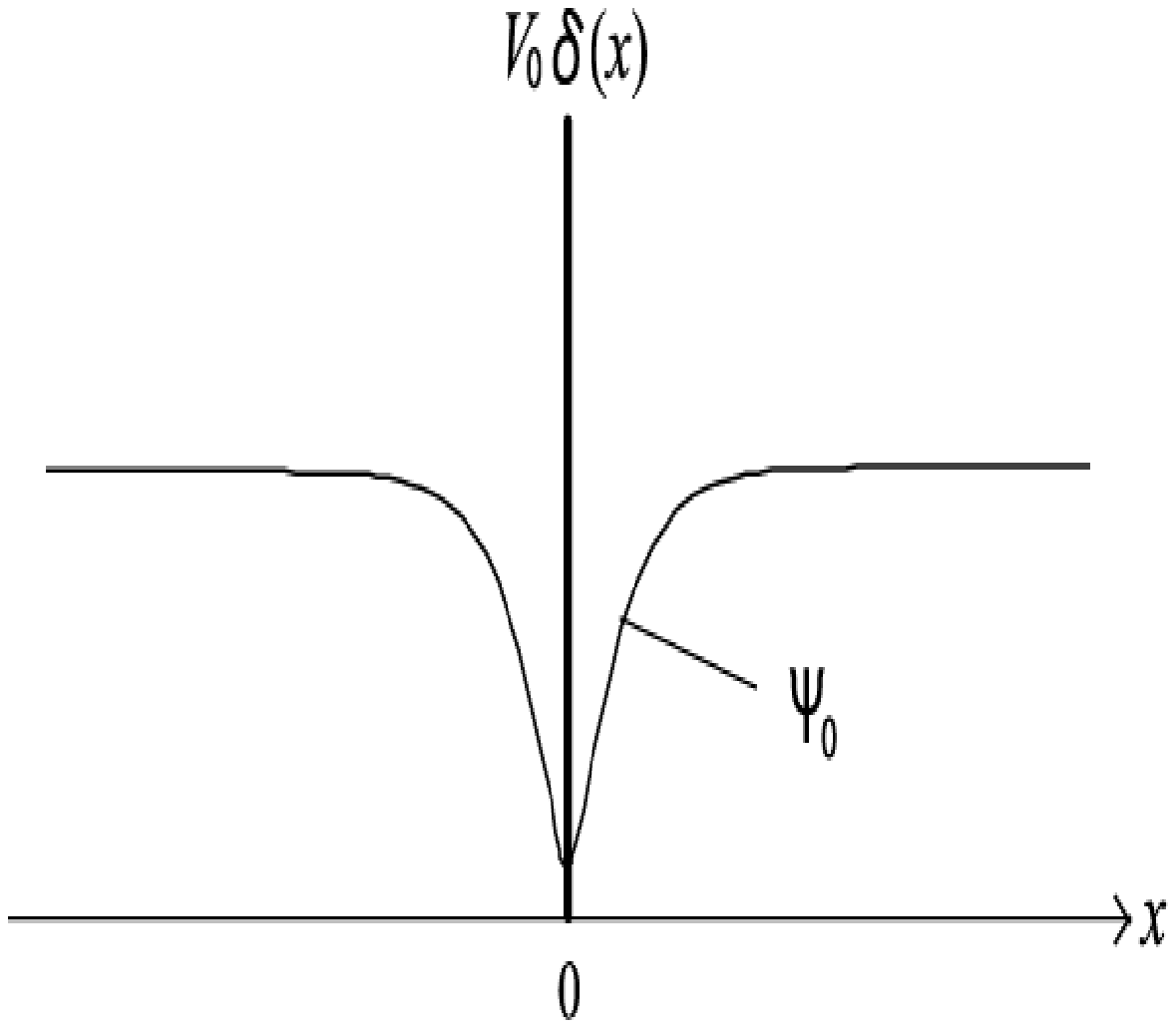}
\caption{\label{fig:tanh}
Schematic picture of the bosonic S-I-S junction.
}
\end{center}
\end{figure}

We consider a BEC at the absolute zero of temperature in a box-shaped trap which consists of a radial harmonic confinement and end caps in the axial direction.
We assume that the frequency $\omega_{\perp}$ of the radial harmonic potential is large enough compared to the excitation energy for the axial direction.
Then, one can justify the one-dimensional treatment of the problem.
Such a configuration was realized in a recent experiment~\cite{rf:box}.
It is assumed that the axial size $L$ of the system is so large that the effect of the edge of the system can be neglected.
Setting a potential barrier at the center of the BEC, one can create a bosonic S-I-S junction.
In order to treat the problem analytically, we adopt a $\delta$-function potential barrier as the potential barrier,
  \begin{eqnarray}
  V(x)=V_0\delta(x).\label{eq:delta}
  \end{eqnarray}
The schematic picture of the bosonic S-I-S junction is shown in Fig. \ref{fig:tanh}.

Our formulation of the problem is based on the mean field theory, which consists of time-independent Gross-Pitaevskii equation and the Bogoliubov equations.
They are written in the dimensionless form as

  \begin{eqnarray}
  [-\frac{1}{2}\frac{d^2}{d\bar{x}^2}-\bar{\mu}+\bar{V}(\bar{x})
  +|\bar{\Psi}_0(\bar{x})|^2] \bar{\Psi}_0(\bar{x}) =0,\label{eq:sGPE}
   \end{eqnarray}
and
       \begin{eqnarray}
               \left(
                 \begin{array}{cc}
                 \bar{H_0} & -{\bar{\Psi}_0(\bar{x})}^2 \\
                 {\bar{\Psi}_0(\bar{x})}^{\ast2} & -\bar{H_0}
                 \end{array}
               \right) 
               \left(
                 \begin{array}{cc}
                 \bar{u}(\bar{x}) \\ \bar{v}(\bar{x})
                 \end{array}
               \right)
               = \bar{\varepsilon}\left(
                 \begin{array}{cc}
                   \bar{u}(\bar{x}) \\ \bar{v}(\bar{x})
                 \end{array}
               \right),  \label{eq:BdGE}\\
               \bar{H_0} = -\frac{1}{2}\frac{d^2}{d\bar{x}^2}
               -\bar{\mu}+\bar{V}\left(\bar{x}\right)+2|\bar{\Psi}_0|^2.
       \end{eqnarray}
We have introduced the following notation:
  \begin{eqnarray}
  \bar{x}=\frac{x}{\xi},\\
  (\bar{\Psi}_0,\bar{u},\bar{v})=\sqrt{\frac{1}{n_0}}(\Psi_0,u,v),\\
  \bar{\varepsilon}=\frac{\varepsilon}{gn_0},\\
  \bar{\mu}=\frac{\mu}{gn_0},\\
  \bar{V}(\bar{x})=\bar{V_0}\,\delta(\bar{x})
       =\frac{V_0}{gn_0}\,\delta(\frac{x}{\xi}),
       \label{eq:dmls}
     \end{eqnarray}
where $\mu$ is the chemical potential, $n_0$ is the density of the condensate fraction for $x\gg\xi$, and $\varepsilon$ is the excitation energy.
The healing length $\xi$ is expressed as $\xi=\frac{\hbar}{\sqrt{mgn_0}}$.
Here, $\Psi_0(x)$ is the wave function of the condensate and $(u(x), v(x))^{\bf t}$ is the wave function of the excitation.
The coupling constant $g$ is affected by the harmonic oscillator length $a_{\perp}$ of the radial confinement as $g=\frac{2\hbar^2a_0}{ma_{\perp}^2}$~\cite{rf:oned}.
We shall omit the bars for all variables in Eqs. (\ref{eq:sGPE})-(\ref{eq:dmls}) hereafter.
It is clear in Eq. (\ref{eq:BdGE}) that the density of the condensates acts as a kind of potential for excitations.

Since the purpose of our study is to investigate the relation between phase difference and the tunneling of elementary excitations, we need to obtain the condensate wave function with phase difference from Eq. (\ref{eq:sGPE}).
It corresponds to a solution of Eq. (\ref{eq:sGPE}) which has stationary Josephson current.
Let us first find such a solution assuming
  \begin{eqnarray}
  \Psi_0(x)={\rm e}^{{\rm i}(qx+C_{\pm})}\label{eq:fref}
  \end{eqnarray}
at far from the potential barrier $|x|\gg1$.
The constant $C_{\pm}$ expresses the phase of the condensate wave function at $x\rightarrow \pm \infty$.
If there is no interatomic interaction, or $g=0$, the solution of Eq. (\ref{eq:sGPE}) satisfying the boundary condition (\ref{eq:fref}) does not exist.
This is because finite fraction of the incident wave is inevitably reflected by the potential barrier.
On the other hand, if there is repulsive interatomic interaction, the solution with the boundary condition (\ref{eq:fref}) exists as shown in Refs.~\cite{rf:hakim, rf:seaman}.
That means that BECs with repulsive interaction go through the potential barrier without reflection, and this behavior clearly exhibits the superfluidity of the BECs.
In this case, the chemical potential is expressed as
  \begin{eqnarray}
  \mu=1+\frac{q^2}{2}.
  \end{eqnarray}
One can realize such a situation in experiments by moving the potential barrier at the velocity of $-\frac{\hbar \,q}{m}$ and choosing the coordinate system with their origin at the potential barrier.
Substituting $\Psi_0=A(x){\rm e}^{{\rm i}S(x)}$, we can rewrite Eq. (\ref{eq:sGPE}) as
   \begin{eqnarray}
      -\frac{1}{2}\left(\frac{d^2A}{dx^2}-q^2A^{-3}\right)
      +(V(x)-1-\frac{q^2}{2})A+A^3\!=\!0,      \label{eq:ampl}
   \end{eqnarray}
   \begin{eqnarray}
      A^2\frac{dS}{dx}=q,\label{eq:conti}
   \end{eqnarray}
where $q$ expresses the current of the condensate fraction in dimensionless form and Eq. (\ref{eq:conti}) corresponds to the equation of continuity.
In dimensional form, the current is expressed as $\frac{\hbar\, q\, n_0}{m}$.
We do not explicitly write $V(x)$ in Eq. (\ref{eq:ampl}) hereafter, because the effect of the $\delta$-function potential barrier appears only in the boundary condition at $x=0$.
Multiplying Eq. (\ref{eq:ampl}) by $\frac{dA}{dx}$, one can integrate the equation as
  \begin{eqnarray}
  -\frac{1}{4}\left(\frac{dA}{dx}\right)^2
  -\frac{q^2}{4}A^{-2}-(\frac{1}{2}+\frac{q^2}{4})A^2+\frac{1}{4}A^4
  =C,\label{eq:midw}
  \end{eqnarray}
where $C$ is an integration constant.
The integration constant can be determined by the boundary conditions
  \begin{eqnarray}
  A(\pm\infty)&=&1,\\
  \frac{dA}{dx}\biggl|\biggr._{\pm\infty}&=&0.
  \end{eqnarray}
Then, Eq. (\ref{eq:midw}) can be written as
  \begin{eqnarray}
  \left(A\frac{dA}{dx}\right)^2=(1-A^2)^2(A^2-q^2).\label{eq:foo}
  \end{eqnarray}
Integrating Eq. (\ref{eq:foo}) again, one obtains
 \begin{eqnarray}
   A^2=\gamma(x)^2+q^2,\label{eq:dens}
 \end{eqnarray}
where
  \begin{eqnarray}
  \gamma(x)\equiv\sqrt{1-q^2}\,\mathrm{tanh}\left(\sqrt{1-q^2}(|x|+x_0)\right).
  \end{eqnarray}
Substituting Eq. (\ref{eq:dens}) into Eq. (\ref{eq:conti}), one can obtain the phase $S(x)$ of the condensate wave function,
  \begin{eqnarray}
   \!\!\!\!\!\!\!S(x)-S(0)&=&\int_{0}^{x}dx\, \frac{q}{A^2}
   \nonumber\\
   &=&qx+{\rm sgn}(x) \Biggl[{\rm tan}^{-1}\left(\frac{\gamma(x)}{q}\right)
   -\mathrm{tan}^{-1}\left(\frac{\gamma(0)}{q}\right)\Biggr].
   \label{eq:phase}
  \end{eqnarray}
One can easily find from Eqs. (\ref{eq:dens}) and (\ref{eq:phase}) that the condensate wave function is expressed as
  \begin{eqnarray}
  \Psi_0(x)={\rm e}^{{\rm i}\left(qx-{\rm sgn}(x)\,\theta_0\right)}
            \left(\gamma(x)-{\rm sgn}(x)\,{\rm i}q\right),
  \label{eq:con_cur}
  \end{eqnarray}
where
  \begin{eqnarray}
  {\rm e}^{{\rm i}\theta_0}\equiv
  \frac{\gamma(0)-{\rm i}q}{\sqrt{\gamma(0)^2+q^2}}.
  \end{eqnarray}
This solution is almost the same as that for a gray soliton~\cite{rf:soliton}, which is a kind of nonlinear excitations of BECs with repulsive interaction.
The only difference from the solution for a gray soliton is the constant $x_0$ in $\gamma(x)$ which depends on the potential strength $V_0$.
The constant $x_0$ can be determined by the boundary conditions at $x=0$,
  \begin{eqnarray}
  \Psi_0(+0)&=&\Psi_0(-0),\\
  \frac{d\Psi_0}{dx}\biggl|\biggr._{+0}
  &=&\frac{d\Psi_0}{dx}\biggl|\biggr._{-0}+2V_0\Psi_0(0).
  \label{eq:con_bnd}
  \end{eqnarray}
Substituting Eq. (\ref{eq:con_cur}) into Eq. (\ref{eq:con_bnd}), one obtains the equation to determine $x_0$,
  \begin{eqnarray}
  \gamma(0)^3+V_0\gamma(0)^2-(1-q^2)\gamma(0)+V_0q^2=0.\label{eq:gamma}
  \end{eqnarray}
Solving Eq. (\ref{eq:gamma}), one obtains $x_0$ as a function of $q$ and $V_0$.
The solution (\ref{eq:con_cur}) has been obtained in Ref.~\cite{rf:hakim}.

We shall next define the phase difference $\varphi$ and express the current $q$ and the constant $x_0$ (or $\gamma(0)$) by using $\varphi$.
According to Ref.~\cite{rf:gl}, the phase difference is given by
  \begin{eqnarray}
  \varphi\equiv q\int_{-\infty}^{\infty}dx\left(\frac{1}{A^2}
            -1\right).\label{eq:phd}
  \end{eqnarray}
Substituting Eq. (\ref{eq:phase}) into Eq. (\ref{eq:phd}), one obtains
  \begin{eqnarray}
  \varphi=2\left[{\rm tan}^{-1}\left(\frac{\sqrt{1-q^2}}{q}\right)
   -\mathrm{tan}^{-1}\left(\frac{\gamma(0)}{q}\right)\right].
   \label{eq:fai}
  \end{eqnarray}
Equations (\ref{eq:gamma}) and (\ref{eq:fai}) yield $q$ and $\gamma(0)$ as functions of $V_0$ and $\varphi$.
Assuming $V_0\gg 1$, we expand Eqs. (\ref{eq:gamma}) and (\ref{eq:fai}) into power series of $\frac{1}{V_0}$, and obtain
  \begin{eqnarray}
  q\simeq \frac{{\rm sin}\varphi}{2V_0}\left(1+\frac{{\rm cos}\varphi}{V_0}
  -\frac{2+{\rm cos}\varphi-3{\rm cos}^2\varphi}{2V_0^2}\right),
  \label{eq:curph}\\
  \gamma(0)\simeq \frac{1+{\rm cos}\varphi}{2V_0}
  \left(\!1\!-\!\frac{1\!-\!{\rm cos}\varphi}{V_0}
  -\frac{1+4{\rm cos}\varphi-3{\rm cos}^2\varphi}{2V_0^2}
  \right) ,\label{eq:gamph}
  \end{eqnarray}
The leading term of Eq. (\ref{eq:curph}) is the well-known relation between the Josephson current and the phase difference.
It is obvious from Eq. (\ref{eq:curph}) that there is the critical current,
  \begin{eqnarray}
  q_{\rm c} \simeq \frac{1}{2V_0}-\frac{1}{4V_0^3}.\label{eq:cricurr}
  \end{eqnarray}
at the critical phase difference,
  \begin{eqnarray}
  \varphi_{\rm c} \simeq \frac{\pi}{2}-\frac{1}{V_0}+\frac{1}{2V_0^2}.
  \end{eqnarray}
As the potential strength increases, the critical current decreases.
The leading term in Eq. (\ref{eq:cricurr}) is consistent with the result in Ref.~\cite{rf:hakim}.
We can easily see from Eqs. (\ref{eq:dmls}) and (\ref{eq:cricurr}) that the critical current $q_{\rm c}$ equals to zero when $g=0$.

\section{Calculations and results}
In this section, we shall solve the Bogoliubov equations with the condensate wave function of Eq. (\ref{eq:con_cur}) which has the phase difference and discuss the tunneling of the elementary excitations.

\subsection{Definition of transmission and reflection coefficients}
We need to clarify the general definition of transmission and reflection coefficients, before we discuss the tunneling of the elementary excitations.
In the case of the tunneling of a single particle, it is well-known that one can easily define the transmission and reflection coefficients by means of the equation of continuity for the probability current.
When one discusses the tunneling of a single particle through a potential barrier $V_{\rm pb}(x)$, one usually considers a situation in which the particle comes from the left (or right).
In this situation, one assumes that the solution $\phi_{\rm sp}(x)$ of the Schr\"odinger equation takes the form of
  \begin{eqnarray}
  \phi_{\rm sp}(x)&=&\left\{\begin{array}{ll}
  \,\,\,\,{\rm e}^{{\rm i}k_{-}x}+a_{\rm sp}\,{\rm e}^{-{\rm i}k_{-}x}, 
  \,\, x\rightarrow -\infty,\\
        \,\,\,\,\,\,\,\,b_{\rm sp}\,{\rm e}^{{\rm i}k_{+}x},\,\,\,\,\,
  x\rightarrow \infty,
  \end{array}\right.
  \end{eqnarray}
where the wave number $k_{\pm}$ is related to the energy of the particle $E$ as
  \begin{eqnarray}
  k_{\pm}\equiv \sqrt{\frac{2m}{\hbar^2}(E-V_{\rm pb}(\pm\infty))}.
  \end{eqnarray}
The equation of continuity for the probability current gives the relation:
  \begin{eqnarray}
  |a_{\rm sp}|^2+\frac{k_{+}}{k_{-}}|b_{\rm sp}|^2=1.
  \end{eqnarray}
This equation means that the transmission coefficient $T_{\rm sp}$ and the reflection coefficient $R_{\rm sp}$ are defined as
  \begin{eqnarray}
  T_{\rm sp}=\frac{k_{+}}{k_{-}}|b_{\rm sp}|^2, \,\, R_{\rm sp}=|a_{\rm sp}|^2.
  \end{eqnarray}
When $V_{\rm sp}(-\infty)=V_{\rm sp}(\infty)$, $T_{\rm sp}$ and $R_{\rm sp}$ coincide with $|b_{\rm sp}|^2$ and $|a_{\rm sp}|^2$, respectively.

We shall define the transmission coefficient $T$ and the reflection coefficient $R$ for the elementary excitations.
Assuming that an excitation comes from the left, we write the asymptotic form of the wave function of the excitation as
  \begin{eqnarray}
{\psi}^l(x)\!\!&=&\!\!
  \left(\!\begin{array}{cc}
           u^l(x) \\ v^l(x)
             \end{array}\!\right)\nonumber\\
  \!\!&=&\!\!
  \left\{\begin{array}{ll}
   \left(\!\begin{array}{cc}
    u_{k_1} \\ v_{k_1}
    \end{array}\!\right){\rm e}^{{\rm i} k_1 x}
          +a^l\left(\!\begin{array}{cc}
                    u_{k_2} \\ v_{k_2}
                         \end{array}\!\right){\rm e}^{{\rm i} k_2 x},
                          x\rightarrow -\infty,\\
          c^l\left(\!\begin{array}{cc}
                    u_{k_1} \\ v_{k_1}
                         \end{array}\!\right){\rm e}^{{\rm i} k_1 x},
                          x\rightarrow \infty,
          \end{array}\right.\label{eq:asym_ex}
  \end{eqnarray}
where $k_{1}$ and $k_{2}$ are real, and they satisfy
  \begin{eqnarray}
  \varepsilon=qk+\sqrt{\frac{k^2}{2}(\frac{k^2}{2}+2)}.\label{eq:bg_sp}
  \end{eqnarray}
Equation (\ref{eq:bg_sp}) is the Bogoliubov excitation spectrum in a uniform system where the condensate has supercurrent proportional to $q$~\cite{rf:fetter}.
Since Eq. (\ref{eq:bg_sp}) is a fourth order equation for $k$, one can solve it and obtain $k_1$ and $k_2$ analytically.
However, since we focus on low energy regions where a specific tunneling behavior appears, we only write the approximate forms of $k_1$ and $k_2$ for $\varepsilon \ll 1$ here,
  \begin{eqnarray}
  k_1 \simeq \frac{\varepsilon}{1+q}-\frac{\varepsilon^3}{8(1+q)^4},
  \label{eq:k1}\\
  k_2 \simeq -\frac{\varepsilon}{1-q}+\frac{\varepsilon^3}{8(1-q)^4}.
  \label{eq:k2}
  \end{eqnarray}
The amplitudes $u_{k}$ and $v_{k}$ are given by
  \begin{eqnarray}
  u_k=\sqrt{\frac{1+\frac{k^2}{2}+\varepsilon-qk}{2(\varepsilon-qk)}}
      {\rm e}^{{\rm i}(qx+{\rm sgn}(x)\frac{\varphi}{2})},\\
  v_k=\sqrt{\frac{1+\frac{k^2}{2}-\varepsilon+qk}{2(\varepsilon-qk)}}
      {\rm e}^{-{\rm i}(qx+{\rm sgn}(x)\frac{\varphi}{2})}.
  \end{eqnarray}

The Wronskian defined as
  \begin{eqnarray}
    \!W\!(\psi^{j\ast}\!, \!\psi^i)\!=\!
    u^{j\ast}\frac{d}{dx}u^{i}\!-\!u^i\frac{d}{dx}u^{j\ast}\!
    +\!v^{j\ast}\frac{d}{dx}v^i\!-\!v^i\frac{d}{dx}v^{j\ast}
  \end{eqnarray}
yields a relation between $a^l$ and $c^l$, which defines the transmission and reflection coefficients.
One can easily prove from the Bogoliubov equations that $W$ is independent of $x$ when $\psi^{j}$ and $\psi^{i}$ have the same energy.
By evaluating $W(\psi^{l\ast}, \psi^l)$, one obtains
  \begin{eqnarray}
  \frac{-k_2(|u_{k_2}|^2+|v_{k_2}|^2)-q}{k_1(|u_{k_1}|^2+|v_{k_1}|^2)+q}|a^l|^2
  +|c^l|^2=1,\label{eq:prob_con}
  \end{eqnarray}
which expresses the conservation law of the energy flux~\cite{rf:antun}.
It is obvious that the reflection coefficient $R$ and transmission coefficient $T$ are defined from Eq. (\ref{eq:prob_con}) as
  \begin{eqnarray}
  R&=&\frac{-k_2(|u_{k_2}|^2+|v_{k_2}|^2)-q}
         {k_1(|u_{k_1}|^2+|v_{k_1}|^2)+q}|a^l|^2,\\
  T&=&|c^l|^2.
  \end{eqnarray}
When there is no supercurrent, $R$ and $T$ are simply given by $|a^l|^2$ and $|c^l|^2$, respectively.

\subsection{Calculation of transmission coefficient}
The condensate wave function of Eq. (\ref{eq:con_cur}) takes the same form as that for a gray soliton except for the constant $x_0$.
Consequently, the solutions of the Bogoliubov equations are also the same as those for a gray soliton which was obtained in Ref.~\cite{rf:soliton}.
Substituting Eq. (\ref{eq:con_cur}) into the Bogoliubov equations, one can analytically obtain four particular solutions.
They are
 \begin{eqnarray}
  u_n(x) &=&
  \Lambda_n {\rm e}^{{\rm i}[(k_n+q) x+{\rm sgn}(x)\frac{\varphi}{2}]}
     \Biggl\{\Biggr.
     \left(1+\frac{k_n^2}{2\varepsilon}\right)\gamma(x)-{\rm i}\,{\rm sgn}(x)
       \nonumber \\
  && \times\left[q+\frac{k_n}{2\varepsilon}(1-q^2-\gamma(x)^2+\varepsilon)
     +\frac{k_n^3}{4\varepsilon}\right]
     \Biggl.\Biggr\},\label{eq:anslt}
                      \\
  v_n(x) &=& 
  \Lambda_n {\rm e}^{{\rm i}[(k_n-q) x-{\rm sgn}(x)\frac{\varphi}{2}]}
     \Biggl\{\Biggr.
     \left(1-\frac{k_n^2}{2\varepsilon}\right)\gamma(x)+{\rm i}\,{\rm sgn}(x)
       \nonumber \\
  && \times\left[q+\frac{k_n}{2\varepsilon}(1-q^2-\gamma(x)^2-\varepsilon)
     +\frac{k_n^3}{4\varepsilon}\right]
     \Biggl.\Biggr\},
      \end{eqnarray}
where $k_n$ satisfies Eq. (\ref{eq:bg_sp}). Approximate forms of $k_3$ and $k_4$ for $\varepsilon \ll 1$ are given by
  \begin{eqnarray}
  k_{3,4}&\simeq& \mp 2 {\rm i}\sqrt{1-q^2}+\frac{q\varepsilon}{1-q^2}
  \mp {\rm i}\frac{(1+2q^2)\varepsilon^2}{4(1-q^2)^{\frac{5}{2}}}\nonumber\\
  &&-\frac{(q+q^3)\varepsilon^3}{2(1-q^2)^4}.\label{eq:k34}
  \end{eqnarray}
Equations (\ref{eq:k1}) and (\ref{eq:k2}) show that $\left(u_1(x), v_1(x)\right)^{\bf t}$ and $\left(u_2(x), v_2(x)\right)^{\bf t}$ describe scattering components.
It is noted that there exist the scattering components whose energy is lower than the condensate potential, because the excitation spectrum is gapless in contrast to the case of the superconductor.
Equation (\ref{eq:k34}) shows that $\left(u_3(x), v_3(x)\right)^{\bf t}$ on the left hand side of the barrier and $\left(u_4(x), v_4(x)\right)^{\bf t}$ on the right hand side describe the localized components around the potential barrier, and $\left(u_4(x), v_4(x)\right)^{\bf t}$ on the left hand side of the barrier and $\left(u_3(x), v_3(x)\right)^{\bf t}$ on the right hand side describe the divergent component far from the potential barrier.
The normalization constant $\Lambda_n$ is expressed as
  \begin{eqnarray}
  \Lambda_n &=& \left\{\!\begin{array}{cc}
  \frac{{\rm e}^{{\rm i}\alpha_1}}{\sqrt{2(\varepsilon-qk_1)}}\,\,\,n=1, 3, 4\\
  \frac{{\rm e}^{{\rm i}\alpha_2}}{\sqrt{2(\varepsilon-qk_2)}}, \,\,\, n=2
                  \end{array}\!\right.,
  \end{eqnarray}
where
  \begin{eqnarray}
  \!\!\!\!\!\!\!\!\!\!\!\!\!\!\!\!\!\!\!\!\!\!\!\!\!
  {\rm e}^{{\rm i}\alpha_n}\equiv
  \frac{4\varepsilon+2\varepsilon qk_n+2(1-q^2)k_n^2+qk_n^3}
  {4\varepsilon\sqrt{1+\frac{k_n^2}{2}+\varepsilon-qk_n}}
  +{\rm i}\,{\rm sgn}(x)\frac{\sqrt{1-q^2}k_n(2\varepsilon-2qk_n+k_n^2)}
  {4\varepsilon\sqrt{1+\frac{k_n^2}{2}+\varepsilon-qk_n}}.
  \end{eqnarray}
The normalization constant of the scattering components is determined to satisfy Eq. (\ref{eq:asym_ex}).

Two independent eigenfunctions of Eq. (\ref{eq:BdGE}) corresponding to two types of scattering process are obtained by omitting divergent components. One is the process in which an excitation comes from the left-hand side (${\psi}^l(x)$), and the other from the right-hand side (${\psi}^r(x)$).
Here we consider the former written as
  \begin{eqnarray}
   \psi^l(x)
   =
     \left(\!\begin{array}{cc}
                      u^l \\ v^l
                  \end{array}\!\right)
   =\left\{\begin{array}{ll}
          \left(\!\begin{array}{cc}
                    u_1 \\ v_1
                         \end{array}\!\right)
          +a^l\left(\!\begin{array}{cc}
                    u_2 \\ v_2
                         \end{array}\!\right)
          +b^l\left(\!\begin{array}{cc}
                    u_3 \\ v_3
                         \end{array}\!\right),
                         & x<0,  \\
          c^l\left(\!\begin{array}{cc}
                    u_1 \\ v_1
                         \end{array}\!\right)
          +d^l\left(\!\begin{array}{cc}
                    u_4 \\ v_4
                         \end{array}\!\right),
                         & x>0,
          \end{array}\right. \label{eq:lcs}
\end{eqnarray}
The coefficients $a^l$, $b^l$, $c^l$, and $d^l$ are the amplitudes of the reflected, the left localized, the transmitted, and the right localized components, respectively. They are functions of the energy $\varepsilon$, the potential strength $V_0$ and the phase difference $\varphi$.
The boundary conditions at $x=0$ yield four equations to determine all the coefficients
  \begin{eqnarray}
     {\psi}^l(+0)&=&{\psi}^l(-0),\\
     \left. \frac{d{\psi}^l}{dx} \right|_{+0}\!\!\!
     &=&\left. \frac{d{\psi}^l}{dx} \right|_{-0}+2V_0{\psi}^l(0)\!\!\!. 
     \label{eq:bc}
  \end{eqnarray}
These equations are linear simultaneous equations for the coefficients $a^l$, $b^l$, $c^l$ and $d^l$, and one can analytically solve them.
Assuming $\varepsilon \ll 1$ and $V_0 \gg 1$, we can obtain analytical solutions of them within the first order of $\varepsilon$ or $\frac{1}{V_0}$, 
  \begin{eqnarray}
  \!\!\!\!\!\!\!\!\!\!\!\!\!\!\!\!\!\!\!\!
  a^l =
          \left[V_0\varepsilon+\frac{(3{\rm cos}\varphi+{\rm sin}\varphi+4)
          \varepsilon}{2}-{\rm i}V_0\varepsilon^2\right]/z^l,
               \\
  \!\!\!\!\!\!\!\!\!\!\!\!\!\!\!\!\!\!\!\!
  b^l = \varepsilon\biggl[\biggr.-{\rm cos}\varphi
          +\frac{-4{\rm cos}^2\varphi+{\rm sin}\varphi\,{\rm cos}\varphi
          -2{\rm cos}\varphi+2}{2V_0}
          +{\rm i}({\rm cos}\varphi+{\rm sin}\varphi)\varepsilon
          \biggr]/(2z^l),
          \\
  \!\!\!\!\!\!\!\!\!\!\!\!\!\!\!\!\!\!\!\!
  c^l = \left[ {\rm i}\!\left(\mathrm{cos}\varphi
         +\frac{3{\rm cos}^2\varphi \!+\! 2{\rm cos}\varphi \!-\!1}{V_0}\right)
           \!+\!2{\rm sin}\varphi\,\varepsilon\right]\!/\! z^l,
           \label{eq:cof_t}\\
  \!\!\!\!\!\!\!\!\!\!\!\!\!\!\!\!\!\!\!\!
  d^l = \varepsilon\biggl[\biggr.{\rm cos}\varphi
          +\frac{4{\rm cos}^2\varphi-{\rm sin}\varphi\,{\rm cos}\varphi
          +2{\rm cos}\varphi-2}{2V_0}
          -{\rm i}\,{\rm cos}\varphi\,\varepsilon
          \biggr]/(2z^l),\\
  \!\!\!\!\!\!\!\!\!\!\!\!\!\!\!\!\!\!\!\!
  z^l = V_0\varepsilon+\frac{(3{\rm cos}\varphi+{\rm sin}\varphi+4)
          \varepsilon}{2}
         +{\rm i}\!\left(\mathrm{cos}\varphi+\frac{3{\rm cos}^2\varphi 
         + 2{\rm cos}\varphi \!-\!1}{V_0}\right).
         \label{eq:cof_de}
  \end{eqnarray}
The relations between the coefficients in ${\psi}^l$ and those in ${\psi}^r$ are
  \begin{eqnarray}
     \begin{array}{cc}
        a^l(\varepsilon,-\varphi)=a^r(\varepsilon,\varphi),\;
        b^l(\varepsilon,-\varphi)=b^r(\varepsilon,\varphi),
         \\
        c^l(\varepsilon,-\varphi)=c^r(\varepsilon,\varphi),\;
        d^l(\varepsilon,-\varphi)=d^r(\varepsilon,\varphi).
     \end{array}
  \end{eqnarray}

When $\varphi$ is not equal to $\varphi_{\rm c}$, we easily obtain the transmission coefficient $T=|c^l|^2$ from Eqs. (\ref{eq:cof_t}) and (\ref{eq:cof_de}).
It is expressed as
  \begin{eqnarray}
  T=\frac{\Delta^2}{\Delta^2+\varepsilon^2},
  \end{eqnarray}
where
  \begin{eqnarray}
  \Delta=\sqrt{\frac{{\rm cos}^2\varphi}{V_0^2}
          +\frac{3{\rm cos}^3\varphi-2{\rm cos}\varphi}{V_0^3}}.
          \label{eq:peakw}
  \end{eqnarray}
The transmission coefficient has a peak at $\varepsilon=0$, and the peak has Lorentzian shape with half width $\Delta$.
This means that the potential barrier is transparent for the low-energy excitations.
This behavior of the transmission coefficient, called {\it anomalous tunneling}, has been predicted for a current-free condensate~\cite{rf:antun,rf:kov}.
We can see from Eq. (\ref{eq:peakw}) that the peak width decreases as $\varphi$ approaches $\varphi_{\rm c}\simeq \frac{\pi}{2}$, and it becomes infinitesimal for $\varphi\rightarrow \varphi_{\rm c}$.
Thus, the anomalous tunneling is suppressed by the supercurrent.
When $\varphi$ reaches $\varphi_{\rm c}$, the transmission coefficient is expressed as
  \begin{eqnarray}
  T\simeq \frac{4}{V_0^2}.
  \end{eqnarray}
Obviously, the anomalous tunneling behavior does not exist any longer, and the transmission coefficient has only a small residual value.

\begin{figure}[p]
\begin{center}
      \includegraphics[width=4.5 in, height=7 in]{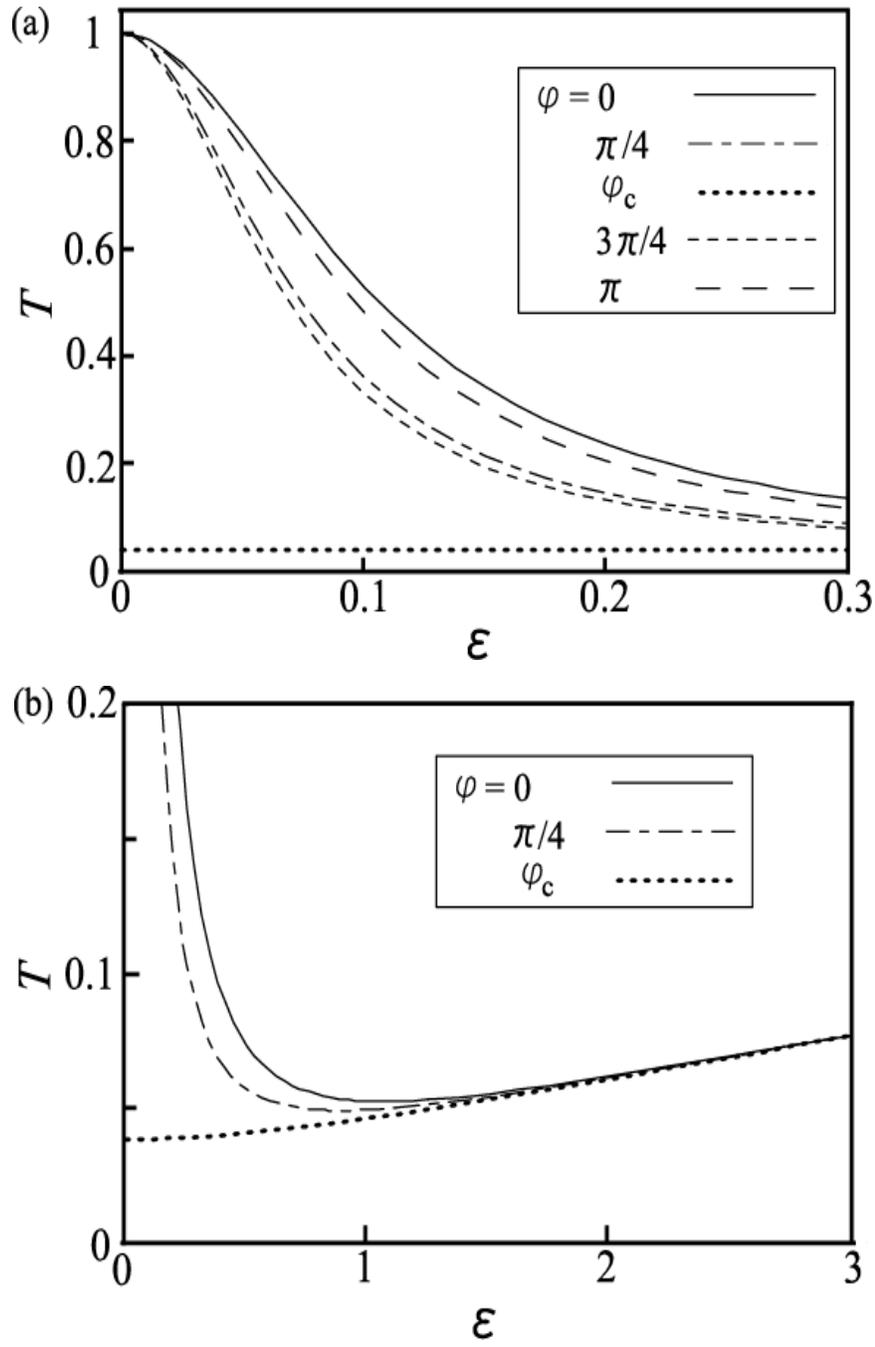}
\caption{\label{fig:antun}
The transmission coefficient $T$ as a function of $\varepsilon$ and $\varphi$, where $V_0=10$.
(a) $T$s at low energy with $\varphi=0$, $\varphi=\pi/4$, $\varphi=\varphi_{\rm c}\simeq\pi/2$, $\varphi=3\pi/4$ and $\varphi=\pi$ are shown.
(b) $T$s are shown up to a higher energy region.
                 }
\end{center}
\end{figure}

We plot the transmission coefficient as a function of $\varepsilon$ at $\varphi=0, \frac{\pi}{4}, \varphi_{\rm c}(\simeq \frac{\pi}{2}), \frac{3\pi}{4}$ and $\pi$ are shown in Fig. 2\label{fig:tau2}, where we set $V_0=10$.
In Fig. \ref{fig:antun}(a) we can clearly see the properties of the transmission coefficient mentioned above.
In the region far from the anomalous tunneling peak, $T$ increases with $\varepsilon$ in a conventional way and its $\varphi$-dependence disappears as shown in Fig. \ref{fig:antun}(b).
This is because the condensate potential terms in the Bogoliubov equations, written as $|\Psi_0|^2$, $\Psi_0^2$ and $\Psi_0^{\ast2}$, begin to become so small compared to the kinetic energy term that the Bogoliubov excitations behave as single particle excitation.

While we adopted the $\delta$-function potential barrier of Eq. (\ref{eq:delta}) in the above calculations, the width of a potential barrier is finite in real systems.
However, the treatment of the problem using the $\delta$-function potential barrier gives us valuable qualitative insights, because it enables us to analytically calculate physical quantities such as the transmission coefficient.
In fact, the result of the calculation of the transmission coefficient for the $\delta$-function potential barrier~\cite{rf:kov} is qualitatively the same as that for the rectangular potential barrier~\cite{rf:antun} in the case of current-free condensate.
Moreover, comparing Eq. (\ref{eq:curph}) to Eq. (\ref{eq:ph_cr}) in Appendix, we can see that the relation between the supercurrent and the phase difference for the $\delta$-function potential is also qualitatively the same as that for the rectangular potential.
Accordingly, we consider that the $\varphi$-dependence of the transmission coefficient is valid not only for the $\delta$-function potential barrier but also for a potential barrier with finite width.

\section{Discussion}
\subsection{Origin of the anomalous tunneling}
Since the anomalous tunneling exhibits behaviors specific to the resonant tunneling, it is expected that its origin can be attributed to the appearance of quasi-bound states with lifetime $\tau_{\rm qb}\sim \frac{\hbar}{\Delta}$.
Kagan {\it et al}. actually insisted that the quasi-bound states were induced by spatial changes of the condensate density.
Their argument is as follows.
The condensate density acts as a kind of potential for excitations.
It inevitably reduces near the potential
 barrier, and consequently creates a potential well for excitations as shown in Fig. \ref{fig:tanh}.
Using an analogy from the resonant tunneling of single particles, such a potential well induces the quasi-bound state, and this is the origin of the anomalous tunneling.


Here we try to elucidate mechanisms of the anomalous tunneling from another viewpoint of the prominence of the localized components with imaginary momenta of Eq. (\ref{eq:k34}).
Calculation of the probability density of each component in the left-incident state $\psi^l$ yields useful information to such viewpoints.
Since the normalization condition for $(u(x),v(x))^{\bf t}$ is given by~\cite{rf:BEC}
   \begin{eqnarray}
      \frac{1}{L}\int dx(|u(x)|^2-|v(x)|^2)=1,
   \end{eqnarray}
the probability density of each component is defined as $|u_1(x)|^2-|v_1(x)|^2$ for the incident component, $|a^l|^2(|u_2(x)|^2-|v_2(x)|^2)$ for the reflected component, $|b^l|^2(|u_3(x)|^2-|v_3(x)|^2)$ for the left localized component, $|c^l|^2(|u_1(x)|^2-|v_1(x)|^2)$ for the transmitted component or $|d^l|^2(|u_4(x)|^2-|v_4(x)|^2)$ for the right localized component.
When $V_0 \gg 1$ and $\varepsilon\ll 1$, one obtains the probability density at $x=0$ of each component,
  \begin{eqnarray}
  |u_1(-0)|^2-|v_1(-0)|^2 \sim \frac{1}{2}+\frac{{\rm sin}\varphi}{2V_0},
  \label{eq:probinci}\\
  |a^l|^2(|u_2(-0)|^2-|v_2(-0)|^2) \sim
  |a^l|^2 (\frac{1}{2}-\frac{{\rm sin}\varphi}{2V_0}),\\
  |b^l|^2(|u_3(-0)|^2-|v_3(-0)|^2) \sim
  \frac{|b^l|^2}{\varepsilon^2}
  \biggl(\biggr.-2-
  \frac{3\!+\!3\,{\rm cos}\varphi+{\rm sin}\varphi}{V_0}
  \biggl.\biggr),\\
  |c^l|^2(|u_1(+0)|^2-|v_1(+0)|^2) \sim
  |c^l|^2 (\frac{1}{2}+\frac{{\rm sin}\varphi}{2V_0}),
  \label{eq:probTC}\\
  |d^l|^2(|u_4(+0)|^2-|v_4(+0)|^2) \sim
  \frac{|d^l|^2}{\varepsilon^2}
  \biggl(\biggr.-2-
  \frac{3+3\,{\rm cos}\varphi+{\rm sin}\varphi}{V_0}
  \biggl.\biggr).\label{eq:}
  \end{eqnarray}

\begin{figure}[p]
\begin{center}
      \includegraphics[width=4.5 in, height=7 in]{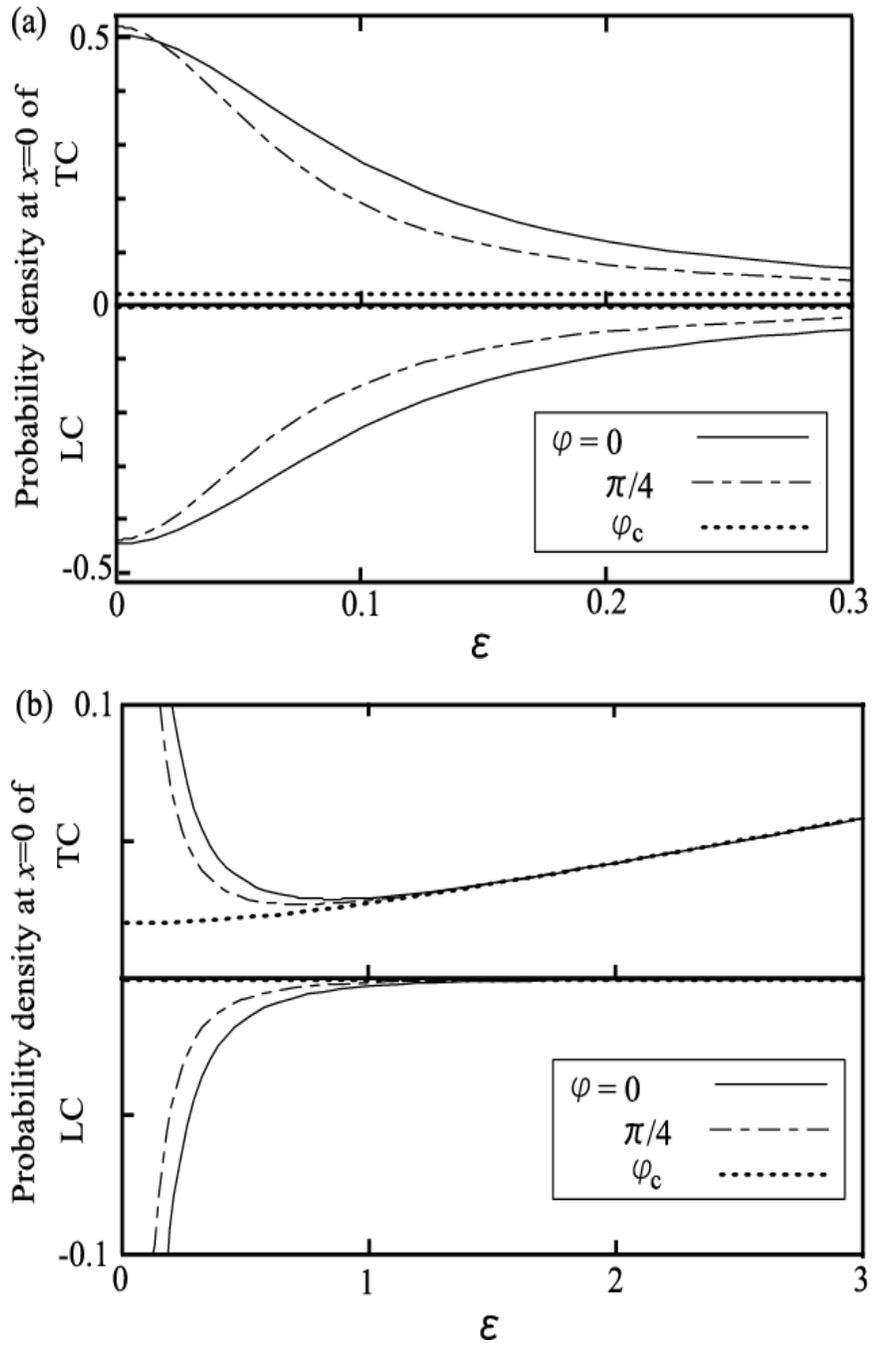}
\caption{\label{fig:pro}
The probability densities at $x=0$ of the the transmitted component (TC) and the right localized component (LC) as a function of 
            $\varepsilon$ and $\varphi$, where $V_0=10$.
            (a) They are shown in a low energy region. 
            (b) They are shown up to a higher energy region. 
                 }
\end{center}
\end{figure}

We show the probability densities at $x=0$ of the transmitted component (TC) and the localized component (LC) in Fig. \ref{fig:pro}.
We can see from Eq. (\ref{eq:probTC}) and Fig. \ref{fig:pro}(a) that the maximum value at $\varepsilon=0$ of the probability density of the transmitted component increases slightly as the phase difference increases.
However, the maximum of the transmission coefficient does not change, because the probability density of the incident component also increases slightly as seen in Eq. (\ref{eq:probinci}).
When $\varphi$ is not equal to $\varphi_{\rm c}$, the localized components prominently appear around $\varepsilon=0$ where the anomalous tunneling occurs.
When $\varphi$ is equal to $\varphi_{\rm c}$, the localized components do not appear.
Thus, the localized components are induced only when the anomalous tunneling is effective.
This result suggests that the localized components are crucial to understand the anomalous tunneling.
It is to be noted that the probability densities of the localized components are negative~\cite{rf:local}.
Due to their negativity, the probability density of scattering components can be comparably large even near the potential barrier without totally raising the probability density of $\psi^l(0)$.
Accordingly, the localized components have a role to spread the scattering components across the potential barrier, and the appearance of the localized component is one of the origins of the anomalous tunneling.

\subsection{Observation of the anomalous tunneling}
In order to describe realistic parameters properly, we add the bars to the dimensionless parameters again as in the Eq. (\ref{eq:dmls}).
Since the width of the potential barrier is finite in real systems, we consider a rectangular potential barrier,
  \begin{eqnarray}
  V(x)=V\theta(\frac{d}{2}-|x|),\label{eq:rec_pote}
  \end{eqnarray}
where $V$ and $d$ are the height and the width of the potential barrier, respectively.
The anomalous tunneling is predicted also for this rectangular potential barrier~\cite{rf:antun}.
When the barrier is so high that $\frac{1}{\kappa_0\xi}, {\rm e}^{-\kappa_0 d}\ll 1$ is satisfied, the peak width of the transmission coefficient can be expressed as~\cite{rf:cross}
  \begin{eqnarray}
  \frac{\Delta_{\rm rec}}{gn_0}\simeq
  \frac{2\sqrt{2}{\rm e}^{-\kappa_0 d}}{\kappa_0\xi},\label{eq:haba}
  \end{eqnarray}
where
  \begin{eqnarray}
  \kappa_0=\sqrt{\frac{2m}{\hbar^2}(V-\mu)}.
  \end{eqnarray}
One needs to create excitations with energy comparable to $\Delta_{\rm rec}$ for the observation of the anomalous tunneling in experiments.
For that reason, the values of two parameters with length dimension are restricted.
One is the barrier width $d$, which corresponds to thickness of the laser beam.
The other is the system size, which is the axial length $L$ of the system.
The healing length $\xi$ and the barrier intensity $\kappa_0\xi$ determine the restrictions.

We consider the experimental setup of Ref.~\cite{rf:box}.
The healing length is $\xi \sim 1 \mu{\rm m}$ in that experiment, because the total number of $^{87}{\rm Rb}$ atoms, the system size and the radial harmonic trap frequency are $N\sim 3000$, $L\sim 80\mu{\rm m}$ and $2\pi\omega_{\perp}\sim 40 {\rm kHz}$.
We see from Eq. (\ref{eq:haba}) that the peak width decreases as the barrier width increases.
This means that the potential barrier should be as narrow as possible.
Since one can narrow down the laser beam waist, which corresponds to the barrier width, to the value comparable to the wavelength, the laser width can be $d\sim1 \mu{\rm m}$ experimentally with barriers created by a blue detuned laser.

We shall give the restriction on the system size.
The anomalous tunneling occurs in a low energy region where the Bogoliubov spectrum is phononlike.
Phonons with energy comparable to or smaller than $\Delta_{\rm rec}$ are available if the axial size of the system $L$ is larger than the wavelength $\lambda_{\Delta}$, where $\lambda_{\Delta}$ is easily calculated from Eq. (\ref{eq:haba}) as $\frac{\lambda_{\Delta}}{\xi} \sim \kappa_0\xi {\rm e}^{\kappa_0 d}$.
We set $d = 1 \,\mu {\rm m}$ and $\xi = 1 \,\mu {\rm m}$.
With a barrier height of $\kappa_0\xi=3$, $L$ should be comparable to or larger than $60 \;\mu\mathrm{m}$.
Since the system size is $L\simeq 70\mu{\rm m}$ in the experimental setup of Ref.~\cite{rf:box}, they barely overcomes this restriction.
Moreover, one can enlarge the system size by changing the position of the optical end caps.
Thus, the anomalous tunneling is expected to occur in the uniform BEC if one can use a sufficiently narrow laser beam to create the potential barrier.

A procedure to observe the anomalous tunneling is as follows.
We consider two BECs separated by a potential barrier.
One can control the phase difference by moving the potential barrier at the velocity $-\frac{\hbar\, q}{m}$, where $q$ is given by Eq. (\ref{eq:ph_cr}) (see Appendix).
At first, one stimulates the condensate on the left hand side into phonon excitations with energy comparable or smaller than $\Delta_{\rm rec}$ by using the Bragg pulse.
The phonon with wavelength $\lambda$ takes $\frac{\lambda}{c_{\rm s}}$ to pass through the barrier, where $c_{\rm s}\equiv \sqrt{\frac{gn_0}{m}}$ is the Bogoliubov sound speed.
After longer time than $\frac{\lambda_{\Delta}}{c_{\rm s}}$, we detect the number of transmitted excitations $n_{\rm t}$ and that of reflected excitations $n_{\rm r}$ by the time of flight absorption images, and the transmission coefficient is given by $T=n_{\rm t}/(n_{\rm t}+n_{\rm r})$.
Thus, the observation of the anomalous tunneling may be possible in an optimized setup.

The dramatic $\varphi$-dependence of the transmission coefficient discussed in Sec. III enables us to determine or estimate the phase difference between two condensates by measuring the transmission coefficient.
The phase difference is near $n\pi$ if most excitations at low energy transmit across the potential barrier, while it is near $(n+\frac{1}{2})\pi$ if most excitations at low energy are reflected by the potential barrier,
where $n$ is an arbitrary integer.
\section{CONCLUSION}
In summary, we have investigated $\varphi$-dependence of the tunneling of the elementary excitations in a bosonic S-I-S junction by solving Bogoliubov equations, and we have found the significant $\varphi$-dependence of the transmission coefficient.

Calculating the probability density of the localized components, we have discussed mechanisms of the anomalous tunneling.
We have shown that the localized component arises only when the anomalous tunneling appears; this means that the prominence of the localized components is one of the origins of the anomalous tunneling.

We have discussed the feasibility of observing the anomalous tunneling in experiments.
We expect that the anomalous tunneling may be observed in BECs trapped in a box-shaped potential.


\ack
The authors would like to thank T. Kimura, S. Takei, S. Tsuchiya and T. Nikuni for fruitful comments and discussions.
They also greatly acknowledge useful comments of D. L. Kovrizhin.
The work is partly supported by a Grant for The 21st Century COE Program (Holistic Research and Education center for Physics  of Self-organization 
Systems) at Waseda University from the Ministry of Education, Sports, Culture, Science and Technology of Japan.
I. D. is supported by JSPS (Japan Society for the Promotion of Science) Research Fellowship for Young Scientists.
\appendix
\section*{Appendix: Josephson relation for a rectangular potential barrier}
\setcounter{section}{1}
We have already obtained the relation between the supercurrent of the condensate and the phase difference for a strong $\delta$-function potential barrier given by Eq. (\ref{eq:curph}), and we have shown that the relation satisfies the Josephson relation.
In this appendix, we shall calculate the relation between the supercurrent of the condensate and the phase difference for a rectangular potential barrier, and show that the Josephson relation is satisfied also in this case.

We consider a BEC in the box-shaped trap mentioned in Sec. II.
Instead of a $\delta$-function potential barrier, we adopt a rectangular potential barrier with width $d$ expressed in the dimensionless form as
  \begin{equation}
  V(x)=V\theta(\frac{d}{2}-|x|).
  \end{equation}
We shall solve Eqs. (\ref{eq:ampl}) and (\ref{eq:conti}) for this potential barrier and obtain the relation between $q$ and $\varphi$.

Outside the barrier, one obtains the solution of the same form as Eqs. (\ref{eq:dens}) and (\ref{eq:phase}),
  \begin{equation}
   A^2=\tilde{\gamma}(x)^2+q^2,\label{eq:dens_rec}
  \end{equation}
  \begin{eqnarray}
   \!\!\!\!\!\!\!\!\!S(x)-S(\pm\frac{d}{2})&=&
   \int_{\pm\frac{d}{2}}^{x}dx\, \frac{q}{A^2}
   \nonumber\\
   &=&q(x\mp\frac{d}{2})+{\rm sgn}(x) 
   \Biggl[{\rm tan}^{-1}\left(\frac{\tilde{\gamma}(x)}{q}\right)
   -{\rm tan}^{-1}\left(\frac{\tilde{\gamma}(\pm\frac{d}{2})}{q}\right)\Biggr],
   \label{eq:phase_rec}
  \end{eqnarray}
where
  \begin{equation}
  \tilde{\gamma}(x)\equiv\sqrt{1-q^2}\,
  {\rm tanh}\left(\sqrt{1-q^2}(|x|-\frac{d}{2}+x_0)\right).
  \end{equation}
Under the barrier, the general solution takes the form~\cite{rf:gl}
 \begin{equation}
  A^2=
  a^2+\frac{{\rm sn}^2(\sqrt{\beta_{+}}x, \sigma)}
  {{\rm cn}^2(\sqrt{\beta_{+}}x, \sigma)}\beta_{-}.
 \end{equation}
where
  \begin{eqnarray}
   a = A(0),\\
  \sigma^2=\frac{\beta_{+}-\beta_{-}}{\beta_{+}},\\
  \beta_{\pm}=
  \frac{3a^2+\kappa_0^2 \pm \sqrt{(\kappa_0^2+a^2)^2-\frac{4q^2}{a^2}}\;}{2},
  \kappa_0=\sqrt{2(V-\mu)}.
  \end{eqnarray}
Here ${\rm cn}(u, q)$ and ${\rm sn}(u, q)$ are the Jacobi elliptic functions.
Assuming $\kappa_0, e^{\kappa_0 d} \gg 1$, one can neglect the last term of Eq. (\ref{eq:ampl}).
In this case, one finds the solution
 \begin{eqnarray}
  A^2=\frac{a^2+\left(\frac{q}{\kappa_0 a}\right)^2}{2}
          {\rm cosh}2\kappa_0 x+
          \frac{a^2-\left(\frac{q}{\kappa_0 a}\right)^2}{2},\\
  S\left(\frac{d}{2}\right)-S\left(-\frac{d}{2}\right)
  =\int_{-\frac{d}{2}}^{\frac{d}{2}}dx\, 
  \frac{q}{A^2}\nonumber\\
  \,\,\,\,\,\,\,\,\,\,\,\,\,\,\,\,\,\,\,\,\,\,\,\,\,\,\,\,\,\,\,\,\,\,\,\,\
  \,\,\,\,\,\,\,\,\,
  \simeq \pi-2\;{\rm tan}^{-1}\left(\frac{\kappa_0 a^2}{q}\right).
   \label{eq:phase_rec2}
 \end{eqnarray}
The constants $x_0$ and $a$ can be determined by the boundary conditions at $x=\frac{d}{2}$,
  \begin{eqnarray}
  \Psi_0\left(\frac{d}{2}+0\right)=\Psi_0\left(\frac{d}{2}-0\right),
  \label{eq:bou_rec}\\
  \frac{d\Psi_0}{dx}\Bigl.\Bigr|_{\frac{d}{2}+0}=
  \frac{d\Psi_0}{dx}\Bigl.\Bigr|_{\frac{d}{2}-0}.\label{eq:boud_rec}
  \end{eqnarray}
Substituting Eqs. (\ref{eq:phase_rec}) and (\ref{eq:phase_rec2}) into Eq. (\ref{eq:phd}), we obtain
  \begin{equation}
  \!\!\!\!\!\!\!\!\!
  \varphi=\pi-2\;{\rm tan}^{-1}\left(\frac{\kappa_0 a^2}{q}\right)+qd
  +2\Biggl[{\rm tan}^{-1}\left(\frac{\tilde{\gamma}(x)}{q}\right)
  -{\rm tan}^{-1}
   \left(\frac{\tilde{\gamma}(\pm\frac{d}{2})}{q}\right)\Biggr],
   \label{eq:phrec}
  \end{equation}
Expanding Eqs. (\ref{eq:bou_rec}), (\ref{eq:boud_rec}) and (\ref{eq:phrec}) into power series of $\kappa_0^{-1}$ and ${\rm e}^{-\kappa_0 d}$, one obtains
  \begin{eqnarray}
  \tilde{\gamma}(\frac{d}{2}) \simeq \frac{1}{\kappa_0},\\
  a \simeq \frac{\sqrt{2(1+{\rm cos}\varphi)}}{\kappa_0}
  {\rm e}^{-\frac{\kappa_0 d}{2}},\\
  q \simeq \frac{2 {\rm e}^{-\kappa_0d}}{\kappa_0}{\rm sin}\varphi.
  \label{eq:ph_cr}
  \end{eqnarray}
We can clearly see from Eq. (\ref{eq:ph_cr}) that the relation between the supercurrent and the phase difference for the rectangular potential barrier takes the form of the Josephson relation as well as the case of the $\delta$-function potential barrier.


\end{document}